\documentclass[fleqn]{article}
\usepackage{amsmath,amsthm,amssymb,bm,lscape}
\begin{document}

\title{Correlation functions for the three state 
superintegrable chiral Potts spin chain of finite lengths}

\author{Klaus Fabricius
\footnote{e-mail Fabricius@theorie.physik.uni-wuppertal.de}\\
Physics Department, University of Wuppertal, 
42097 Wuppertal, Germany\\
Barry~M.~McCoy
\footnote{e-mail mccoy@max2.physics.sunysb.edu}\\               
C.N.Yang Institute for Theoretical Physics,\\ 
State University of New York,
 Stony Brook,  NY 11794-3840}
\date{\today}

\maketitle

\begin{abstract}
We compute the correlation functions of the three state superintegrable chiral
Potts spin chain for chains of length 3,4,5. From these results we
present conjectures for the form of the nearest neighbor correlation function.

\end{abstract}

\section{Introduction}

The free energy of the Ising model was first solved by Onsager \cite{ons} 
in 1944 who invented a
method of solution based on what is now known as ``Onsager's algebra''.
This algebra is generated from  operators $A_0$ and $A_1$ 
which form a Hamiltonian
\begin{equation}
H=A_0+\lambda A_1
\label{ham}
\end{equation}
where
\begin{equation}
[A_0,[A_0,[A_0,A_1]]]=const [A_0,A_1]
\end{equation}
and from $A_0$ and $A_1$ the full algebra is generated as \cite{perk}
\begin{eqnarray}
&&[A_j,A_k]=4G_{j-k}\nonumber\\
&&[G_m,A_l]=2A_{l+m}-2A_{l-m}\nonumber\\
&&[G_j,G_k]=0
\label{onsalg}
\end{eqnarray}
For 41 years the Ising model was the only known model which satisfied
this algebra but in 1985 von Gehlen and Rittenberg \cite{vgr} 
made the remarkable
discovery that the Hamiltonian (\ref{ham}) with 
\begin{eqnarray}
A_0=-\sum_{j=1}^{\mathcal N}\sum_{r=1}^{N-1}
\frac{e^{i\pi(2r-N)/(2N)}}{\sin\pi r/N}Z^r_jZ^{\dagger r}_{j+1} \label{a0}\\
A_1=-\sum_{j=1}^{\mathcal N}\sum_{r=1}^{N-1}
\frac{e^{i\pi(2r-N)/(2N)}}{\sin\pi r/N}X^r_j\label{a1}
\end{eqnarray}
where $Z_j$ and $X_j$  are direct product matrices
\begin{eqnarray}
&&Z_j=I\otimes\cdots\otimes Z\otimes \cdots \otimes I\\
&&X_j=I\otimes\cdots\otimes X\otimes \cdots \otimes I
\end{eqnarray}
$I$ is the $N\times N$ identity matrix, the $N\times N$ matrices
$Z$ and $X$ are in the $j^{th}$ position in the product and have the  
matrix elements
\begin{eqnarray}
&&Z_{j,k}=\omega^{j}\delta_{j,k}\\
&&X_{j,k}=\delta_{j,k+1}
\end{eqnarray}
with
\begin{equation}
\omega=e^{2\pi i/N}
\end{equation}
is also a representation of Onsager's algebra (\ref{onsalg}). The
Hamiltonian (\ref{ham}) with $A_0$ and $A_1$ given by (\ref{a0}) and
(\ref{a1}) is called the $N$ state superintegrable chiral Potts spin chain.
When $N=2$ the Hamiltonian of the superintegrable chiral 
Potts spin chain reduces to the Hamiltonian studied by Onsager \cite{ons}.
 
It was discovered in \cite{ampt} by means of  explicit computations on
chains of small length that 
the eigenvalues of the superintegrable
chiral Potts Hamiltonian are all of the form
 \begin{equation}
E=A+B\lambda+N\sum_{j=1}^m\pm(1+\lambda^2+a_j\lambda)^{1/2}
\label{eform}
\end{equation}
This form was proven to follow directly from Onsager's algebra by
\cite{davies1,davies2,date}. 
The parameters  $a_j,A,B$ in (\ref{eform}) 
 have been computed by the method of functional equations
\cite{bax,amp1,amp2,bbp} where it is found, for $\lambda$ suitably less
than unity, that in the thermodynamic limit the ground state energy
per site is
\begin{equation}
e_0(\lambda)=\lim_{\cal N \rightarrow \infty}
\frac{1}{\mathcal N}E_0(\lambda;{\mathcal N})
=-(1+\lambda)\sum_{l=1}^{N-1}F(-\frac{1}{2},\frac{l}{N};1;
\frac{4\lambda}{(1+\lambda)^2})
\end{equation}
where $F(a,b;c;z)$ is the hypergeometric function.

In this paper we extend
these finite chain computations from the eigenvalues of the
Hamiltonian to the correlation functions in the ground state
$\langle Z^r_0 Z^{\dagger r}_R \rangle$. 

There are several constraints these correlations must satisfy.
One such constraint is the obvious requirement that
\begin{equation}
\langle Z^{(N-r)}_0Z^{\dagger (N-r)}_1\rangle=\langle Z^r_0Z^{\dagger
  r}_1\rangle^*
\label{zzherm}
\end{equation}
Furthermore in thermodynamic limit ${\mathcal N} \rightarrow \infty$ the
correlation is related to the order parameter by 
\begin{equation}
M^2_r=\lim_{R\rightarrow \infty}\langle Z^r_0Z^{\dagger r}_R\rangle
\label{mscpdef}
\end{equation}
For $N=2$ the order parameter is the spontaneous magnetization of
the Ising model
\begin{equation}
M=(1-\lambda^2)^{1/8}
\label{misingresult}
\end{equation}
which was reported by Onsager \cite{ons2} in 1948  and 
proven by Yang \cite{yang} in 1952.
For the $N$ state superintegrable chiral Potts spin chain it was conjectured by
Albertini, McCoy, Perk and Tang \cite{ampt} in 1988 and proven by
Baxter \cite{bax2} in 2005 that 
\begin{equation}
M_r=(1-\lambda^2)^{r(N-r)/2N^2}
\label{mscpresult}
\end{equation}

The order parameter of the Ising model (\ref{misingresult}) may be
computed \cite{mpw} by use of Szeg{\H o}'s theorem applied 
to the representation of the
correlation function $\langle Z_0Z^{\dagger}_R\rangle$ as a 
determinant \cite{ko} which is derived using free fermion methods first
invented by Kaufmann \cite{kauf}. However, the chiral Potts order
parameter (\ref{mscpresult}) is computed by Baxter \cite{bax2} by
functional equation methods which do not extend to a computation of
the correlation functions $\langle Z^r_0Z^{\dagger r}_R\rangle$

Because the superintegrable chiral Potts model is a generalization of
the Ising model with the same underlying Onsager algebra there
must be structure of the Ising correlations which generalizes to
superintegrable chiral Potts. However, because the Ising correlations
have been computed by means of free Fermi methods and not by use of Onsager
algebra this structure remains unknown. 

Recently Au-Yang and Perk \cite{ap1,ap2,ap3,ap4} and Nishino and 
Deguchi \cite{deg} have initiated
the study of the eigenvectors of the superintegrable chiral Potts
spin chain by use of the Onsager algebra. These important studies are
the necessary foundation for the computation of the correlation functions
from the point of view of the Onsager algebra. 

The purpose of this paper is to provide insight into the correlation
functions of the superintegrable chiral Potts spin chain by
explicitly calculating the correlations $\langle Z^r_0Z^{\dagger r}_R\rangle$
for the three state case $N=3$ for chains of finite length ${\cal
  N}=3,4,5$ which extends to correlation functions the study of \cite{ampt}
of the ground state energy $E_0(\lambda,{\mathcal N})$. For any value
of ${\mathcal N}$ the nearest neighbor correlations satisfy a sum rule
coming from the ground state energy $E_0(\lambda;{\mathcal N})$. This
sum rule is presented and discussed in sec. 2. In sec. 3 we present
the results of our finite chain computations and we conclude in sec. 4
with a discussion of the implications which the results for finite
chains have for the correlations in the thermodynamic limit.

\section{Sum rule}

There is an elementary result known as Feynman's theorem that for any
Hamiltonian which depends on a parameter $\lambda$ that
\begin{equation}
\frac{\partial}{\partial \lambda}\langle H(\lambda)\rangle=
\frac{\partial}{\partial \lambda}E_0(\lambda)
\label{feyn}
\end{equation}
where $\langle O \rangle$ denotes the expectation value of the
operator $O$ in the ground state and $E_0(\lambda)$ denotes the ground
state energy as a function of $\lambda$.
 Therefore it follows from (\ref{feyn}) that for the superintegrable
 chiral Potts Hamiltonian (\ref{ham}),(\ref{a0}),(\ref{a1}) 
the expectation $\langle X^r_0 \rangle$
satisfies the sum rule
\begin{eqnarray}
&&-{\mathcal N}\sum_{r=1}^{N-1}\frac{e^{i\pi(2r-N)/(2N)}}
{\sin \pi r/N}\langle X_0^r \rangle
=\frac{\partial E_0(\lambda;{\mathcal N})}{\partial \lambda}\nonumber\\
&&=B-N\sum_{j=1}^m\frac{\lambda+a_j/2}{(1+\lambda^2+a_j\lambda)^{1/2}}
\end{eqnarray}
and $\langle Z^r_0Z^{\dagger r}_1\rangle$ satisfies
\begin{eqnarray}
&&-{\mathcal N}\sum_{r=1}^{N-1}\frac{e^{i\pi(2r-N)/(2N)}}{\sin \pi r/N}
\langle Z^r_0Z^{\dagger r}_1\rangle=
E_0(\lambda;{\mathcal N})
-\lambda\frac{\partial E_0(\lambda;{\mathcal N})}{\partial \lambda}\nonumber\\
&&=A-N\sum_{j=1}^m\frac{1+\lambda a_j/2}{(1+\lambda^2+a_j \lambda)^{1/2}}
\label{zzsum}
\end{eqnarray}

\section {Correlations for $N=3$ and ${\mathcal N}=3,4,5$}

We will explicitly consider the three state case $N=3$
where the Hamiltonian (\ref{ham}) is explicitly written as
\begin{equation}
H=-\sum_{j=1}^{\mathcal N}
\left((1-\frac{i}{\sqrt 3})(Z_jZ_{j+1}^{\dagger}+\lambda X_j)
+(1+\frac{i}{\sqrt 3})(Z^2_jZ^{\dagger 2}_{j+1}+\lambda X_j^2)\right)
\end{equation}
and using (\ref{zzherm}) the sum rule (\ref{zzsum}) is
\begin{equation}
(1-\frac{i}{\sqrt 3})\langle Z_0Z_1^{\dagger}\rangle
+(1+\frac{i}{\sqrt 3})\langle Z_0Z_1^{\dagger}\rangle^*
=-\frac{A}{\mathcal N}+\frac{N}{\mathcal N}\sum_{j=1}^m
\frac{1+a_j\lambda/2}{(\lambda^2+1+a_j\lambda)^{1/2}}
\end{equation}
It is
known from \cite{amp2} that for small $\lambda$ the ground state is in
the sector $P=0$ and $Q=0$ where $P$ is the momentum of the state and
\begin{equation}
e^{i 2 \pi Q/3}=\prod_{k=1}^{\mathcal N}X_k
\end{equation}
For the ground state
\begin{equation}
A=B=-P_a,~~{\rm for}~{\mathcal N}\equiv -P_a {\rm mod}~3~~P_a=0,1,2
\end{equation}
and 
\begin{equation}
a_j=\frac{2(1+t^3_j)}{1-t^3_j}
\label{atrel}
\end{equation}
and $t_j$ are the roots of the polynomial equation
\begin{eqnarray}
&&0=t^{-P_a}\{(t-1)^{\mathcal N}(t\omega^2-1)^{\mathcal
    N}\omega^{-P_a}\nonumber\\
&&~~~+(t\omega-1)^{\mathcal N}(t\omega^2-1)^{\mathcal N}  
 +(t-1)^{\mathcal N}(t\omega-1)^{\mathcal N}\omega^{P_a}\}
\label{tequ}
\end{eqnarray}
with $\omega=e^{2\pi i/3}$. From computations of Galois groups done on
Maple we see for ${\mathcal N}\leq 12$ that (\ref{tequ}) can be
explicitly solved in terms of radicals only for ${\mathcal N}=3,4,5,6,9,12$

The eigenspace for the states with $P=Q=0$ are of dimension 5 for
${\mathcal N}=3$, dimension 8 for ${\mathcal N}=4$ and dimension 17
for ${\mathcal N}=5$. The correlations may now be computed in
principle by computing the normalized eigenvectors in the subspace 
$P=Q=0$ and then explicitly computing the matrix elements of
$Z_0Z_R^{\dagger}$. In practice the algebra is too formidable to do by
hand and the computation must be computerized. When this is done and
expressions are simplified by removing common factor the results are as
follows

\subsection{${\mathcal N}=3$} 
From (\ref{atrel}) and (\ref{tequ}) we find
\begin{equation}
9a^2-20=0
\end{equation} 
and thus
\begin{equation}
a_1=-\frac{2\sqrt 5}{3},~~~~~a_2=\frac{2\sqrt 5}{3}
\end{equation}
Thus defining
\begin{equation}
x_j=(\lambda^2+1+a_j\lambda)^{1/2}
\label{xjdef}
\end{equation}
we find
\begin{equation}
\langle Z_0Z^{\dagger}_1\rangle=
(1-i{\sqrt 3})\left(\frac{3}{40}-\frac{3\lambda^2-7}{40x_1x_2}\right)
+\frac{3}{8}(1+\frac{i}{\sqrt 3})
\left(\frac{1+a_1\lambda/2}{x_1}+\frac{1+a_2\lambda/2}{x_2}\right)
\label{result3}
\end{equation}
 From (\ref{result3}) we find
\begin{equation}
(1-\frac{i}{\sqrt 3})\langle Z_0Z^{\dagger}\rangle=
-i\frac{4}{\sqrt 3}\left(\frac{3}{40}-\frac{3\lambda^2-7}{40x_1x_2}\right)
+\frac{1}{2}
\left(\frac{1+a_1\lambda/2}{x_1}+\frac{1+a_2\lambda/2}{x_2}\right)
 \end{equation}
and hence the sum rule (\ref{zzsum}) is obviously satisfied.
For  $\lambda \rightarrow 0$ (\ref{result3}) is expanded as
\begin{equation}
 \langle Z_0Z^{\dagger}_1\rangle=1-\frac{2}{9}\lambda^2
-\frac{17-i3{\sqrt 3}}{54}\lambda^4+O(\lambda^6)
 \end{equation}
and for $\lambda\rightarrow \pm \infty$
\begin{equation}
\langle Z_0Z^{\dagger}_1\rangle=\frac{1}{3}(1+\frac{i}{\sqrt 3})|\lambda|^{-1}
+\frac{1}{6}(1-i{\sqrt 3})\lambda^{-2}+O(|\lambda|^{-3})
\end{equation}

\subsection{${\mathcal N}=4$}
 For ${\mathcal N}=4$ 
\begin{equation}
27a^2+36a-20=0
\end{equation}
and thus there are again two roots $a_j$ which are now
 given by
\begin{equation}
a_1=-\frac{2}{3}+\frac{4}{9}{\sqrt 6},~~~~~
a_2=-\frac{2}{3}-\frac{4}{9}{\sqrt 6}
\end{equation}
We define $x_j$ as before (\ref{xjdef}) and obtain for 
$\langle Z_0Z_1^{\dagger}\rangle$ the result
\begin{eqnarray}
&&\langle Z_0Z^{\dagger}_1\rangle=\frac{1}{4}
-(1-i{\sqrt 3})\frac{2\lambda-3}{40x_1x_2}\nonumber\\
&&+\left(1+a_1\lambda/2\right)
\frac{3}{320 x_1}[36-{\sqrt 6}+i{\sqrt 2}(3+2{\sqrt 6})]\\
&&+\left(1+a_2\lambda/2\right)
\frac{3}{320 x_2}[ 36+{\sqrt 6}-i{\sqrt2}(3-2{\sqrt 6}]\nonumber\\
&&=\frac{3}{16}(1+\frac{i}{\sqrt 3})+\frac{1}{16}(1-i{\sqrt 3})
-(1-i{\sqrt 3})\frac{2\lambda-3}{40x_1x_2}\nonumber\\
&&+\left(1+a_1\lambda/2\right)
\frac{3}{320 x_1}][ 30(1+\frac{i}{\sqrt 3})+(6-{\sqrt 6})(1-i{\sqrt 3})]
\nonumber\\
&&+\left(1+a_2\lambda/2\right)
\frac{3}{320 x_2}[ 30(1+\frac{i}{\sqrt 3})+(6+{\sqrt 6})(1-i{\sqrt 3})]
\label{result4}
\end{eqnarray}
We note that the coefficients of $1+i3^{-1/2}$ and $1-i3^{1/2}$ are
both real, that
\begin{eqnarray}
&&(1-\frac{i}{\sqrt 3})\langle Z_0Z^{\dagger}_1\rangle\nonumber\\ 
&&=\frac{1}{4}-i\frac{1}{4{\sqrt 3}}
+i\frac{1}{\sqrt 3}\frac{2\lambda-3}{10x_1x_2}\nonumber\\
&&+\left(1+a_1\lambda/2\right)
\frac{3}{320 x_1}][40-i(6-{\sqrt 6})\frac{4}{\sqrt 3}]
\nonumber\\
&&+\left(1+a_2\lambda/2\right)
\frac{3}{320 x_2}[ 40-i(6+{\sqrt 6})\frac{4}{\sqrt 3}]
\label{result4a}
\end{eqnarray}
and thus the sum rule (\ref{zzsum}) is satisfied.
For $\lambda\rightarrow 0$ (\ref{result4a}) reduces to
\begin{equation}
\langle Z_0Z_1^{\dagger}\rangle=1-\frac{2}{9}\lambda^2+
\frac{1+i{\sqrt 3}}{162}\lambda^4+\frac{8{\sqrt 3}i}{729}\lambda^5+O(\lambda^6)
\end{equation}
and for $\lambda\rightarrow \pm \infty$
\begin{equation}
\langle Z_0Z_1^{\dagger}\rangle
=\frac{1}{3}(1+\frac{i}{\sqrt 3})|\lambda|^{-1}
+\frac{1}{18}(1-i{\sqrt 3})\lambda^{-2}
+\frac{4}{81}(1-\frac{4i}{\sqrt 3})|\lambda|^{-3}+O(\lambda^{-4}).
\end{equation}

We also compute $\langle Z_0Z^{\dagger}_2\rangle$ and find
\begin{eqnarray}
&&\langle Z_0Z^{\dagger}_2\rangle
=\frac{1}{4}-\frac{2\lambda-3}{10x_1x_2}\nonumber\\
&&+(1+a_1\lambda/2)\frac{3(6-{\sqrt 6})}{80x_1}
+(1+a_2\lambda/2)\frac{3(6+{\sqrt 6})}{80 x_2}
\end{eqnarray}
For small $\lambda$ this reduces to
\begin{equation}
\langle Z_0Z_2^{\dagger}\rangle
=1-\frac{2}{9}\lambda^2-\frac{1}{81}\lambda^4+O(\lambda^5)
\label{result42}
\end{equation}
and for $\lambda \rightarrow \pm \infty|$
\begin{equation}
\langle Z_0Z_2^{\dagger}\rangle
=\frac{2}{9}\lambda^{-2}+\frac{20}{81}|\lambda|^{-3}+\frac{84}{729}\lambda^{-4}
+O(\lambda^{-5})
\end{equation}

\subsection{${\mathcal N}=5$}

For ${\mathcal N}=5$ there are three roots $a_j$ 
that satisfy
\begin{equation}
81a^3+54a^2-228a-88=0
\end{equation}
which are
given by
\begin{eqnarray}
&&a_1=-\frac{2}{9}-\frac{2}{9}
(W_{+}+W_{-})\\
&&a_2=-\frac{2}{9}-\frac{2}{9}
(\omega^2W_{+}+\omega W_{-})\\
&&a_3=-\frac{2}{9}-\frac{2}{9}
(\omega W_{+}+\omega^2 W_{-})
\end{eqnarray}
where 
\begin{equation}
W_{\pm}=\frac{1}{17}(7\pm i{\sqrt 19})\left(\frac{5}{2}\right)^{1/3}w_{\pm}
\end{equation}
with
\begin{equation}
w_{\pm}=(311\pm i9{\sqrt 19})^{1/3}
\end{equation}
We define $x_j$ as before (\ref{xjdef}) and obtain for 
$\langle Z_0Z^{\dagger}_1\rangle$ the result

\begin{eqnarray}
&&\langle Z_0Z_1^{\dagger}\rangle
=\frac{3}{40}(1+\frac{i}{\sqrt 3})+\frac{37}{40\cdot 19}(1-i{\sqrt 3})
+\frac{9}{40}(1+\frac{i}{\sqrt 3})
\sum_{j=1}^3\frac{1+a_j\lambda/2}{x_j}\nonumber\\
&&+\frac{1}{2^3\cdot 5^2 \cdot19}(1-i{\sqrt 3})\sum_{j=1}^3
\frac{91+b_j-27a_j/2-\lambda(67/3+7b_j/3-7\cdot 9a_j/2)}{x_j}\nonumber\\
&&+\frac{1}{2^3\cdot 5^2\cdot 19}(1-i{\sqrt 3})
\sum_{1\leq j <k \leq 3}\frac{P_{j,k}}{x_jx_k}
\label{result5}
\end{eqnarray}
where
\begin{eqnarray}
&&b_1=1+\left(\frac{5}{2}\right)^{1/3}(w_{+}+w_{-})\nonumber\\
&&b_2=1+\left(\frac{5}{2}\right)^{1/3}(\omega^2 w_{+}
+\omega w_{-})\nonumber\\
&&b_3=1+\left(\frac{5}{2}\right)^{1/3}(\omega w_{+}
+\omega^2 w_{-}),\nonumber\\
\end{eqnarray}
and
\begin{eqnarray}
&&P_{j,k}=164+\frac{81}{2}(a_j+a_k)+7(b_j+b_k)\nonumber\\
&&~~~~~+\frac{\lambda}{3}[-56+\frac{49\cdot 9}{2}(a_j+a_k)
+2(b_j+b_k)]\nonumber\\
&&~~~~~+\left(\frac{\lambda}{3}\right)^2[-504-13\cdot 81(a_j+a_k)
-9\cdot 13(b_j+b_k)]\nonumber\\
\end{eqnarray}

It is easy to see that (\ref{result5}) satisfies the sum rule
(\ref{zzsum}). When $\lambda \rightarrow 0$ (\ref{result5}) reduces to
\begin{equation}
\langle Z_0Z_1^{\dagger}\rangle=1-\frac{2}{9}\lambda^2+\frac{1}{162}
(-9+i{\sqrt 3})\lambda^4+\frac{56}{729}\lambda^5 +O(\lambda^6)
\label{smallresult4}
\end{equation}
and for $\lambda \rightarrow \pm \infty$
\begin{equation}
\langle Z_0Z_1^{\dagger}\rangle=\frac{1}{3}(1+\frac{i}{\sqrt 3})|\lambda|^{-1}
+\frac{1}{18}(1-i{\sqrt 3})\lambda^{-2}
+\frac{4}{81}(1+\frac{i}{\sqrt 3})|\lambda|^{-3}+O(\lambda^{-4})
\end{equation}

\section{Discussion}

From the results  (\ref{result3}),(\ref{result4}) and (\ref{result5})
for $N=3$
we conjecture for arbitrary $\mathcal N$ that
$\langle Z_0Z_1^{\dagger}\rangle$ has the form
\begin{equation}
\langle Z_0Z_1^{\dagger}\rangle=P_0+\sum_{j}\frac{P_j(\lambda)}{x_j}
+(1-i{\sqrt 3})\sum_{j<k}\frac{P_{j,k}(\lambda)}{x_jx_k}
\end{equation}
where $P_0$ is a constant, $P_j(\lambda)$ is  a polynomial linear in
$\lambda$ and $P_{j,k}(\lambda)$ is  a polynomial at most quadratic in
$\lambda$ with real coefficients.
In the limit ${\mathcal N}\rightarrow \infty$ the correlation
$\langle Z_0Z_1^{\dagger}\rangle$ will be a double integral. For $N=2$ 
this correlation is a single integral and thus we expect that for
general $N$ the correlation $\langle Z^r_0Z_1^{r\dagger}\rangle$
will be an $N-1$ fold integral.

We expect  for arbitrary $N$ that
$\langle Z^r_0Z^{r\dagger}_R\rangle$ will be a $(N-1)R$ fold integral 
but the first evidence for this for $N=3$ can only come ${\mathcal N}=6$.

We also conjecture from the result (\ref{result42}) for
$\langle Z_0Z^{\dagger}_2\rangle$ and ${\mathcal N}=4$ that
for all even ${\mathcal N}$ the correlation for $N=3$
$\langle Z_0Z^{\dagger}_{{\mathcal N}/2}\rangle$ is real.
In the limit ${\mathcal N} \rightarrow \infty$ we must have
\begin{equation}
\lim_{{\mathcal N} \rightarrow \infty}
\langle Z_0^rZ^{\dagger r}_{{\mathcal N}/2}\rangle=(1-\lambda^2)^{r(N-r)/N^2}
\end{equation}

\vspace{.2in}

{\bf Acknowledgement}

The bulk of this computation was only possible using the program FORM
written by Prof. J.A.M. Vermaseren.

\end{document}